\documentstyle[epsfig,amssymb,preprint,aps,prc]{revtex}

\newcommand\NPA{{Nucl. Phys.} A}

\newcommand\PLB{{Phys. Lett.} B}

\newcommand\PRL{Phys. Rev. Lett.}
\newcommand\PRC{{Phys. Rev.} C}
\newcommand\PRD{{Phys. Rev.} D}

\newcommand\ZPA{{Z. Phys.} A}

\newcommand\EPJC{{Eur. Phys. J.} C}

\newcommand\JPG{{J. Phys.} G}

\newfam\BMath
\font\BMathL=cmmib10 
\font\BMathl=cmmib7
\font\BMathm=cmmib5
\textfont\BMath=\BMathL \scriptfont\BMath=\BMathl
\scriptscriptfont\BMath=\BMathm

\renewcommand\a{\alpha}

\renewcommand\d{\delta}

\newcommand\q{\theta}

\newcommand\p{\pi}
\newcommand\r{\rho}
\newcommand\s{\sigma}

\renewcommand\o{\omega}

\newcommand\D{\Delta}       

\newcommand\ci{{\cal I}}

\newcommand\cs{{\cal S}}

\newcommand\ra{\rightarrow}

\newcommand\lra{\longrightarrow}

\newcommand{\lan}{\langle}     
\newcommand{\ran}{\rangle}     

\newcommand\be{\begin{equation}}
\newcommand\ee{\end{equation}}
\newcommand\bea{\begin{eqnarray}}
\newcommand\eea{\end{eqnarray}}

\newcommand\ba{\begin{array}}
\newcommand\ea{\end{array}}

\newcommand\eref[1]{Eq.~(\ref{#1})}

\newcommand\bfi{\begin{figure}}
\newcommand\efi{\end{figure}}
\newcommand\bpi[1]{\begin{picture}#1}
\newcommand\epi{\end{picture}}

\newcommand{\fref}[1]{Fig.~\ref{#1}}

\newcommand{\tref}[1]{Table~\ref{#1}}

\def\jou#1#2#3#4{{#1} {\bf #2}, #3 (#4)}

\def\qh{\q_{1/2}}
\def\adq{\lan \D Q\ran}
\def\vm{v_{\text{max}}}

\newcommand\M{{\fam\BMath M}}
\newcommand\N{{\fam\BMath n}}
\newcommand\R{{\fam\BMath r}}
\newcommand\V{{\fam\BMath v}}

\begin{document}


\draft

\title{Bremsstrahlung from a Microscopic Model of \\ Relativistic
Heavy Ion Collisions}

\author{S.M.H. Wong$^1$, M. Belkacem$^1$, J.I. Kapusta$^1$, \\
S.A. Bass$^2$, M. Bleicher$^3$\footnote{Feodor Lynen Fellow of the 
Alexander v. Humboldt Foundation}, H. St\"ocker$^4$} 

\address{
$^1$School of Physics and Astronomy, University of Minnesota, 
Minneapolis, MN 55455, USA     \\
$^2$National Superconducting Cyclotron Laboratory and
Department of Physics and Astronomy, Michigan State University,
E. Lansing, MI 48824, USA \\
$^3$Nuclear Science Division, Lawrence Berkeley National Laboratory,
Berkeley, CA 94720, USA \\
$^4$Institut f\"ur Theoretische Physik, Robert-Mayer-Strasse 10, 
Johann Wolfgang Goethe Universit\"at,    \\
D-60054 Frankfurt am Main, Germany 
}

\maketitle

\begin{abstract}
We compute bremsstrahlung arising from the acceleration of individual 
charged baryons and mesons during the time evolution of high-energy Au+Au 
collisions at the Relativistic Heavy Ion Collider using a microscopic 
transport model. We elucidate the connection between bremsstrahlung and 
charge stopping by colliding artificial pure proton on pure neutron nuclei.  
From the intensity of low energy bremsstrahlung, the time scale and the degree 
of stopping could be accurately extracted without measuring any hadronic 
observables. 
\end{abstract}

\vspace*{1cm}

\pacs{PACS: 25.75.-q, 13.85.Qk \hfill NUC-MINN-00/11-T}

\section{Introduction}
\label{s:intro}

The long awaited Relativistic Heavy Ion Collider (RHIC) at 
Brookhaven National Laboratory has now begun operation. 
It is widely expected that heavy ion collisions at RHIC will result 
in the creation of hot and dense deconfined QCD matter, 
termed the Quark-Gluon-Plasma (for a review on the properties and 
signatures of the plasma, we refer to \cite{qm}).
Apart from searching for this plasma, a question of great interest 
is how bulk matter behaves in a very energetic collision.
This is an old question, going back to the 1950's and therefore 
pre-dating QCD by 20 years.  An early hypothesis, due to Landau 
\cite{Landau}, is that matter would come to a complete 
halt before it is driven to explode by the extremely high pressure. 
Another hypothesis is due to Bjorken \cite{Bj} where projectile and target 
pass through each other but leave behind a zone of hot and dense matter.  

To find out whether the initial nuclei are left with none 
of their original velocity or whether they will be traveling with
almost the original velocity at the end of the collision, 
the most obvious method is to measure the final hadron rapidity 
distributions as is done by the various experimental collaborations; 
for example, see \cite{wienold96a,na49}.  
This method is certainly direct and straightforward, but it is 
not the simplest. Indeed, to do this properly it is necessary to measure 
various baryon species in a wide range of acceptance, not an easy task
in a collider environment such as RHIC.
Furthermore, the measurement of the final-state baryons cannot really
distinguish between primordial baryons (initially contained in the two
colliding nuclei), which have undergone massive stopping and newly created
baryons, which are already produced at their final rapidity values.  

Although strong stopping has been observed in collisions of heavy 
nuclei at the GSI/SIS, BNL/AGS and CERN/SPS accelerators 
\cite{wienold96a,na49}, it is by no means clear whether this trend 
will continue up to the far higher available collision energies 
at RHIC (the maximum achievable energy at the SPS is 
$\sqrt{s} = 8.6$ GeV/nucleon for Pb+Pb, whereas RHIC will collide
Au+Au at up to $\sqrt{s}= 100$ GeV/nucleon). The pertinent question 
is what particular measurements will provide information on the 
degree of energy, momentum, baryon, and electric charge stopping. 
The degree of stopping correlates with the maximum energy density 
achieved in the collision, albeit in complicated ways. 

Already more than 20 years ago bremsstrahlung was suggested as a simple means to 
determine the degree of stopping in nucleus-nucleus collisions \cite{kap}.  That 
first paper concerned nuclear collisions at laboratory energies of order 1 GeV 
per nucleon.  This was later followed by theoretical studies at laboratory 
energies of order 100 GeV per nucleon in \cite{BM,Dumitru}.  More recently, very 
soft photons were studied in \cite{jkcs} along with a detector design to measure 
them at RHIC.  This was followed by further theoretical studies in 
\cite{kstalk,kw} where it was shown that the coarse grained space-time evolution 
of central nucleus-nucleus collisions at RHIC could be discerned by measuring 
photons with energies up to 200 MeV.  Bremsstrahlung
exploits the fact that initially the ions are moving 
relativistically and are highly charged (Z=79 for Au). Therefore 
they copiously emit bremsstrahlung when they experience 
any deceleration due to strong interactions. Because of the high energies 
involved, these relativistically moving, highly charged target and 
projectile will radiate bremsstrahlung in the extreme forward or 
backward directions. The intensity is determined mainly by the amount 
of slowing down experienced during the collisions. So by strategically 
placing photon detectors around the beam pipe in a narrow 
cone, the amount of stopping can be determined \cite{kstalk}. 
In addition, the space-time evolution of the colliding matter can be probed 
by this much simpler method \cite{kw}. 

In \cite{kw} and \cite{eesg}, various models for the time evolution of 
heavy ion collisions at RHIC were used. There are also some similar 
earlier works on the subject using numerical dynamical models but at lower 
energies \cite{vmg,sumgv,hmsg}. In this paper a microscopic hadronic 
model, referred to as the Ultra-relativistic Quantum Molecular Dynamics 
model (UrQMD) \cite{urqmd1,urqmd2}, will be used to calculate bremsstrahlung. 
Since this is a dynamical model that follows the space-time evolution of the 
initial target and projectile as well as all the produced hadrons, it 
should allow us to discover what can be expected for bremsstrahlung
from realistic collisions. Being able to follow the detailed collision
history will give us a better insight into what bremsstrahlung can tell us 
about the collisions.  The method of calculation will be detailed in 
Sect. \ref{s:urqcd} and the results presented in Sect. \ref{s:res}.

In \cite{sw} a quantity $\cs$, obtainable both from hadron rapidity 
distributions and from bremsstrahlung measurements, was defined so that 
it is equal to unity when there is full charge stopping and to zero when 
there is total transparency.  We will discuss the extraction of $\cs$
from the microscopic evolution of heavy ion collisions as treated by 
UrQMD. This together with results from UrQMD, will be presented in 
Sect. \ref{s:res} and \ref{s:mtm}.

\section{Ultra-Relativistic Quantum Molecular Dynamics Model}
\label{s:urqcd}

\subsection{The Model}
\label{s:urqcd:ml}

The UrQMD model is a dynamical transport model that follows the time
evolution of a heavy-ion collision in the
entire many-body phase space. UrQMD has been applied to heavy-ion 
collisions in the energy range from a few hundreds of A$\cdot$MeV to 
several hundreds of A$\cdot$GeV using the same basic concepts and
physics inputs at all energies. A detailed description of the
underlying concepts and comparisons to experimental data are available
in Refs. \cite{urqmd1,urqmd2}.

The model includes explicitly 55 different baryon species
(nucleons, deltas, hyperons, and their known resonances \cite{pdb} up 
to masses of 2.25 GeV) and 32 different meson species (including the
known meson resonances \cite{pdb} up to masses of 2 GeV), as well as
their respective anti-particles and all isospin-projected states.
At RHIC energies, the treatment of subhadronic degrees of freedom is 
of great importance. These degrees of freedom enter in UrQMD via 
the introduction of a formation time for hadrons produced in string
fragmentation. Hadrons produced through string decays have a non-zero
formation time, $\tau_f$, which depends on the energy-momentum of the
particle.  Newly formed particles cannot interact during their
formation time.  The leading hadrons interact within their formation 
time with a reduced cross-section, which is taken to be proportional to 
the number of their original constituent quarks.  

All hadrons are propagated (in a relativistic cascade sense)
according to Hamilton's equations of motion, supplemented by a
relativistic Boltzmann-Uehling-Uhlenbeck collision term involving all
hadron states.  The collision term is based on tabulated or
parameterized experimental cross-section (when available). Resonance
absorption and scattering is handled via the principle of detailed
balance.  If no experimental information is available, the 
cross-section is either calculated via a One-Boson-Exchange model or 
via a modified additive quark model. 

\subsection{The Method}
\label{s:urqcd:md}

The formula for computing classical bremsstrahlung from a space and
time-dependent current is well-known \cite{js}. For our purposes, 
it is convenient to express the current 
as a sum of separate currents denoting the flow of individual moving
charges. For a charge $q_i$ with position $\R_i(t)$ and 
velocity $\V_i(t)$ the current is 
\be {\fam\BMath J}_i (\R_i,t) = q_i \V_i(t)\; 
     \delta \left( \R_i-\R_{i}(0)- \int_0^t dt'\, \V_i(t') \right) \; . 
\ee
The intensity and number of photons emitted with frequency $\omega$ in 
the direction $\N$ is 
\be \frac{d^2 I}{d\o d\Omega} = \o \frac{d^2N}{d\o d\Omega} = \o^2\; |{\bf A}|^2
\ee
where the amplitude is
\be {\bf A} = \sum_i \int 
              \frac{dt\; d^3 \R_i}{4{\p}^{3/2}}  \;
         \left( \N\times [\N \times {\fam\BMath J}_i (\R_i,t)] \right) 
              \; e^{i\o ( t-\N \cdot \R_i(t) )}  \; .
\label{eq:ampdef}
\ee 
The sum is over all charged particles present in the system at
any given time $t$. Unlike in \cite{jkcs,kw}, we are using units
with $\a =e^2/4\p=1/137$ which accounts for the odd factor of 
$1/2 \sqrt{\p}$ in \eref{eq:ampdef}. 

For low energy photons the details of the individual hadron collisions, 
such as whether they are elastic or inelastic, are unimportant. 
For most of the hadron collisions and decays of relevance in this study 
the typical proper, physically estimated spatial and temporal extent
is on the order of 1 fm.  Photons with energy and momentum less than about 
200 MeV will not resolve these details.  Consider the following
examples.  Suppose that the center-of-mass
of a set of particles involved in one of the microscopic collisions
or decays moves with speed $v$ along the beam axis with a corresponding
Lorentz factor $\gamma$ relative to the frame in which the bremsstrahlung
is computed, which is the nucleus-nucleus center-of-velocity frame.
If the proper time for this collision or decay is $\tau \approx 1$ fm/c
then in the computational frame it takes a time $\gamma \tau$.  The
ability of a photon to resolve this is determined by the phase in
\eref{eq:ampdef} which is $\omega \gamma \tau (1 - v \cos\theta)$.
A photon with $\omega < 1/\left( \tau \gamma (1-v\cos\theta) \right)$ 
will not be sensitive to the fine details of this process,
only to the gross features: namely, the difference between the
incoming and outgoing currents.  If $v = 0$ then $\omega$ would have
to be comparable to $1/\tau \approx 200$ MeV.  In the other limit,
$\gamma \gg 1$, the radiation is strongly peaked at $\theta \approx
1/(2\gamma)$ \cite{js}, and so $\omega$ would have to be as large as
$\gamma/\tau$ to resolve the details.  These sorts of details are not
desired; a proper description of them would require knowledge
of the dynamics on the level of quarks and gluons, not hadrons.  Since
there is very little radiation away from the forward direction, and
since one does not expect transverse motion of the afore-mentioned
frames of reference with large $\gamma$, the limitation to photons
with energy less than 200 MeV should be safe.      

As a consequence of the above discussion we treat the
accelerations of the participating hadrons as instantaneous.
In fact, that is exactly how UrQMD treats collisions and decays, even 
when string formation and decay is involved.  For example, a string may
decay by emitting a hadron at $({\bf x}_1,t_1)$, then a second 
hadron at $({\bf x}_2,t_2)$, and so on.  Electric charge is conserved
by each of these instantaneous events.  In UrQMD particles created with
high energy have an associated formation time.  Within this time they
are not allowed to interact via the strong interactions.
If that hadron has an electric
charge it does contribute to the electromagnetic radiation amplitude
immediately, otherwise charge conservation would be violated!
We do not justify or criticize the UrQMD approach; rather, our goal
is to point out that bremsstrahlung can be used to test it and other
detailed dynamical models of high energy nuclear collisions.

For each hadron involved in a microscopic event $e$ with space-time
coordinate $({\bf x}_e,t_e)$, whether it be a 
collision or a decay, we write the acceleration as a 
delta function at the time of the event in terms of its 
velocity $\V_e^{\text{in}}$ before and $\V_e^{\text{out}}$ after the event  
\be  {\fam\BMath a}_e(t) = (\V_e^{\text{out}} - \V_e^{\text{in}}) \; 
\d (t-t_e) \, . 
\ee
Equivalently, the velocity can be written as a stepwise function
\be  \V_e(t) = \V_e^{\text{out}} \q (t-t_e) + \V_e^{\text{in}} \q (t_e-t)    \; . 
\ee 
After some manipulation of \eref{eq:ampdef} one gets the contribution
to the full amplitude ${\bf A}$ contributed by event $e$ as 
\bea {\bf A}_e   = \frac{i}{4\p^{3/2}\o}   \;
 e^{ i\o ( t_e-\N\cdot \R_e ) } \sum_j & & q_j 
\left( \frac{(\N\cdot\V_{je}^{\text{out}}) \N-\V_{je}^{\text{out}} }
{ 1-\N\cdot\V_{je}^{\text{out}} }
-\frac{(\N\cdot\V_{je}^{\text{in}}) \N-\V_{je}^{\text{in}} }
{ 1-\N\cdot\V_{je}^{\text{in}} } \right) \nonumber \\
  = \frac{i}{4\p^{3/2}\o}   \; 
 e^{ i\o ( t_e-\N\cdot \R_e ) } \sum_j & &
 q_j \left( \frac{ \N-\V_{je}^{\text{out}} }{ 1-\N\cdot\V_{je}^{\text{out}} }
           -\frac{ \N-\V_{je}^{\text{in}} }{ 1-\N\cdot\V_{je}^{\text{in}} }
             \right )  \; .
\label{eq:amp}
\eea
This expression is valid for both decays and for elastic and inelastic
collisions.  The sum over $j$ is a sum over all hadrons participating
in the event.  If a hadron is created in the event its $\V_e^{\text{in}}$
is set to zero, since that is equivalent to removing it from the
incoming current.  Similarly, a hadron which is annihilated in
the event has its $\V_e^{\text{out}}$ set to zero.
Note also that any hadron whose velocity does not change at $t_e$
gives no net contribution to the amplitude.

We use UrQMD to generate an ensemble of nucleus-nucleus collisions 
with impact parameter $b \leq 3$ fm at RHIC energy of 
$\sqrt{s}=$ 100 GeV/nucleon. For each nucleus-nucleus collision 
we compute the amplitude in the center-of-mometum frame 
by summing over its space-time evolution of 
microscopic scatterings and decays. We have computed 160 events for the 
results in the sections to come. The moduli squared of the individual 
nucleus-nucleus collisions are then averaged over this ensemble of 
events. Coherence is included within a given nucleus-nucleus collision, 
not between different collisions. These data enable one to calculate 
from \eref{eq:amp} the soft radiation amplitude and therefore the 
photon intensity distribution that can reasonably be expected at RHIC.

\section{Results}
\label{s:res}

\subsection{$\q$ and $\o$ dependence}
\label{s:qo}

Using the method described in the previous sections we compute 
bremsstrahlung arising from central Au-Au collisions at RHIC.
The angular dependence of the intensity distribution at $\o=10$ MeV is 
plotted in \fref{f:th}.  As expected at such high beam energies, 
relativistic effects tend to focus the radiation in the extreme forward and 
backward directions. Such focussed emission should facilitate practical 
measurements of photons.  The $\o$ dependence for several different angles 
are plotted in \fref{f:om}.  In \cite{kw} the $\o$ dependence was studied in 
both Landau-like and in Bjorken-like scenarios.  The Bjorken-like scenario 
results in an essentially flat $\o$-distribution, while a Landau-like 
scenario results in an oscillatory behavior.   The oscillatory behavior 
in the latter scenario follows from interference between the stopping phase 
and the re-acceleration phase of the nuclear collision.  \fref{f:om} shows 
Bjorken-like behavior at small angles but Landau-like behavior at larger angles. 
For $\q =10^0$ and $20^0$ there are definitely hints of oscillation. 

\bfi
\centerline{\epsfig{figure=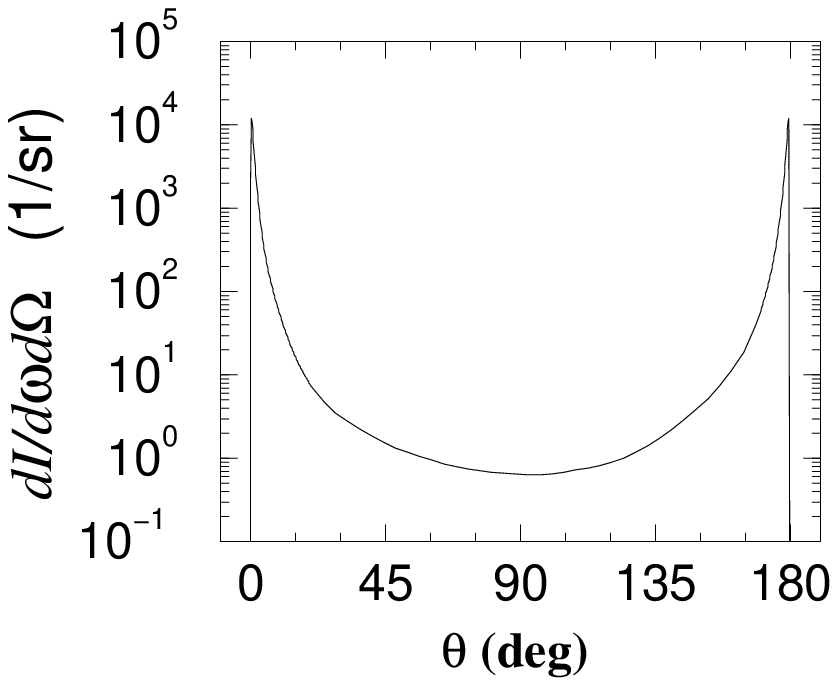,width=3.50in}}
\caption{Angular distribution of the photon intensity from central Au+Au 
collisions at RHIC at a photon energy of 10 MeV.  This shows the expected 
extreme forward-backward focussing.}
\label{f:th}
\efi
\bfi
\centerline{\epsfig{figure=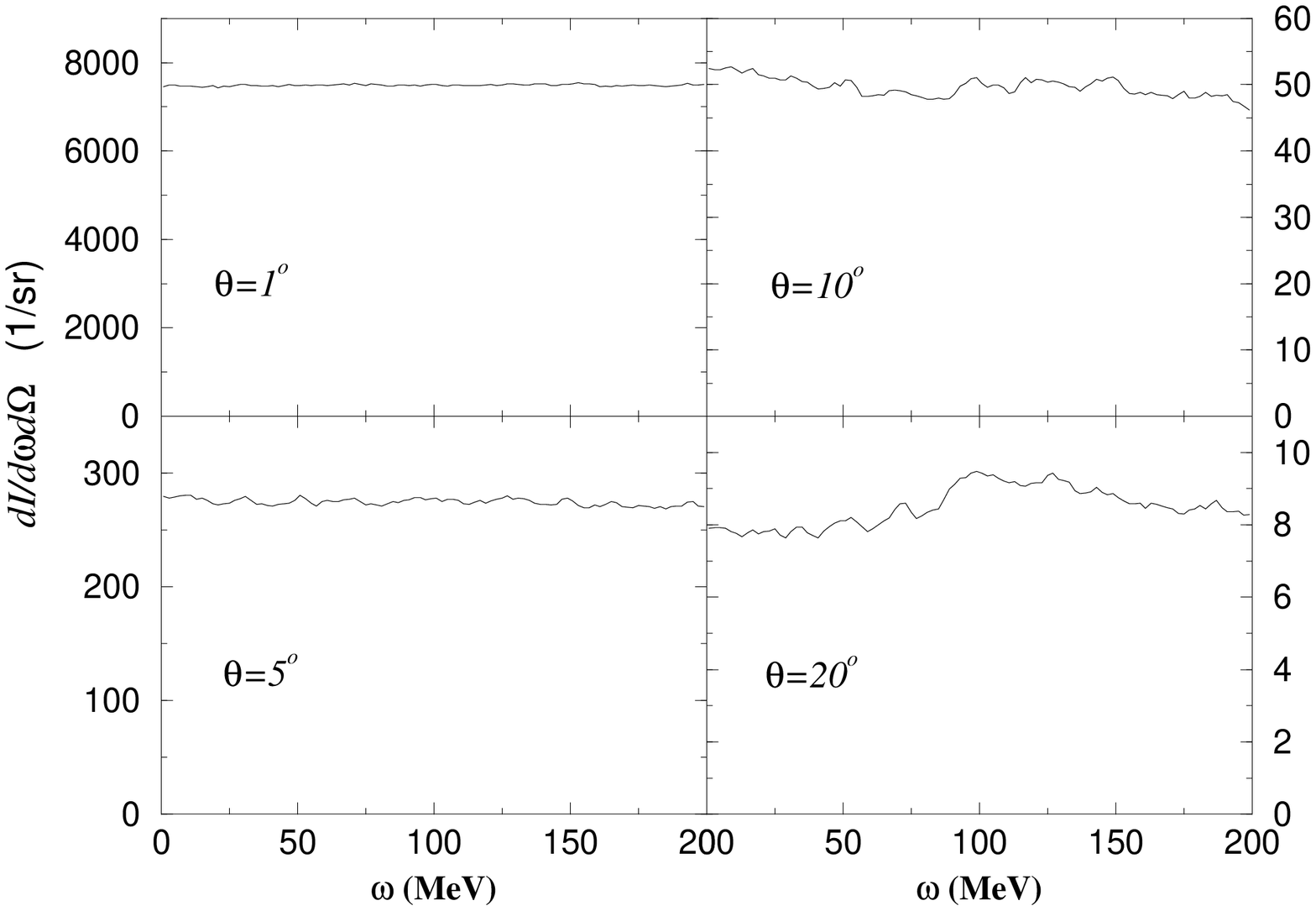,width=6.0in}}
\caption{Photon intensity versus the energy $\o$ of the
photons at fixed angles. The intensity is almost independent of photon energy.
At larger angle $\q$, though, there is a trace of oscillation.} 
\label{f:om}
\efi
\bfi
\centerline{\epsfig{figure=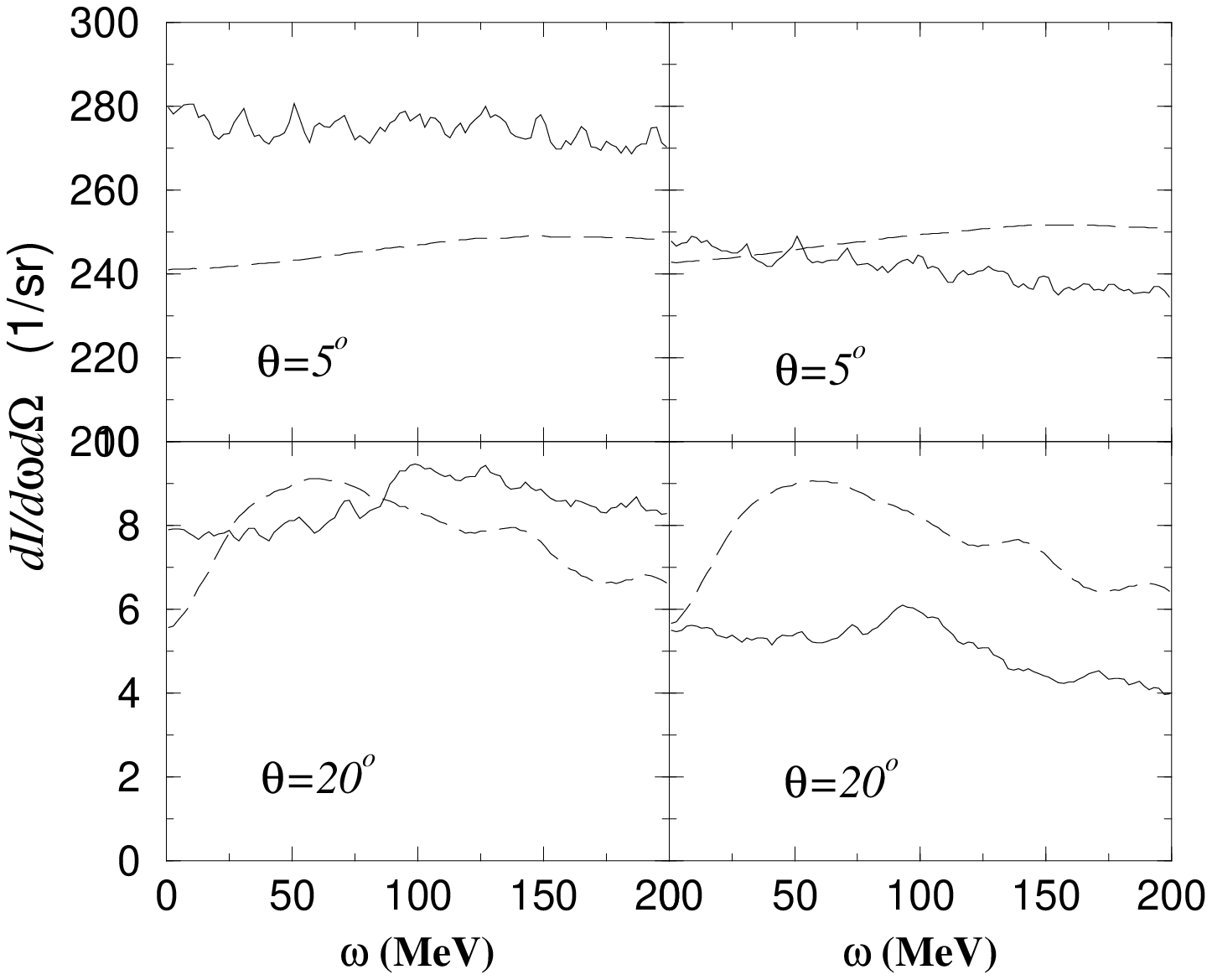,width=6.0in}}
\caption{$\o$-dependence of the bremsstrahlung intensity at two
angles are shown. The solid lines are the total bremsstrahlung and the 
dashed lines are photons from baryons only.  The right panel shows what 
happens when the transverse velocities are arbitrarily set to zero.  The 
baryonic contributions show little dependence
whereas the dependence of the total contributions 
are quite visible. Oscillations in the baryonic component
are clearly apparent.} 
\label{f:nvp}
\efi

The oscillation can be better seen in \fref{f:nvp} where only
$\q=5^o$ and $20^o$ are plotted. The left panels show the total 
bremsstrahlung (solid lines) and the bremsstrahlung originating from 
baryons only (dashed lines). In the right panels, we show the result of
arbitrarily setting the transverse velocities to zero.  It may be seen that 
bremsstrahlung from the baryons is essentially independent of their transverse 
motion. We will make use of this fact in a later section on stopping. 
On the contrary, the charged meson contribution and, therefore,
the total receives a noticeable contribution, especially at larger angles.
Without the mesons it is clearly apparent that there are oscillations from 
the baryons, as pointed out in \cite{kw}.  The relative sensitivity of the 
mesons has two origins.  First, most of the mesonic component comes from 
pions which are very light compared to protons and therefore suffer 
proportionately greater acceleration for the same momentum kick.  Second, 
entropy drives a net increase of protons relative to neutrons, with a 
consequent excess of $\pi^-$ over $\pi^+$ to conserve charge.

\subsection{A=Z=197 on A=N=197}
\label{s:borl}

To elucidate the nature of stopping we have collided an artificial 
nucleus consisting entirely of 197 protons with another artificial one 
consisting entirely of 197 neutrons. Initially, the proton nucleus 
travels in the positive z-direction and the neutron nucleus in the 
negative z-direction. If the final protons all move with positive 
rapidity and the neutrons all with negative rapidity then we clearly have a 
Bjorken-like scenario.  With such an asymmetric charge 
distribution the bremsstrahlung would have a highly asymmetric
$\q$-distribution too.

The angular distribution of 10 MeV photons radiated during central collisions 
of these artificial nuclei, as computed by UrQMD, is plotted in \fref{f:pnth}. 
Instead of two strong peaks in the extreme forward and backward directions, 
as in \fref{f:th}, there is now a strong peak only in the forward direction. 
In the backward direction there is only a weak peak. The reason is that the 
original protons suffer a massive deceleration to rapidities much less than 
their original beam rapidity $y_0$. This is most clearly seen in the net 
nucleon rapidity distributions, $p-\overline{p}$ and $n-\overline{n}$, 
shown in \fref{f:pnrap}. Apart from the remnants of the initial 
(artificial) nuclei appearing as peaks at $y \sim  y_0$ for protons and 
$y \sim - y_0$ for neutrons, the nucleons spread out rather evenly
throughout the entire rapidity range.  Since the elementary nucleon-nucleon 
collisions are known to be strongly forward-backward peaked, this flatness of 
the nucleon distributions in nucleus-nucleus collisions is as likely a result 
of repeated charge exchange as it is from single hard nucleon-nucleon 
scattering or sequential baryon-baryon collisions.  From the point of view 
of electric charge there is no way or means to distinguish.  Our conclusion 
from UrQMD is that electric charge suffers severe deceleration during the 
collision.   

\bfi
\centerline{\epsfig{figure=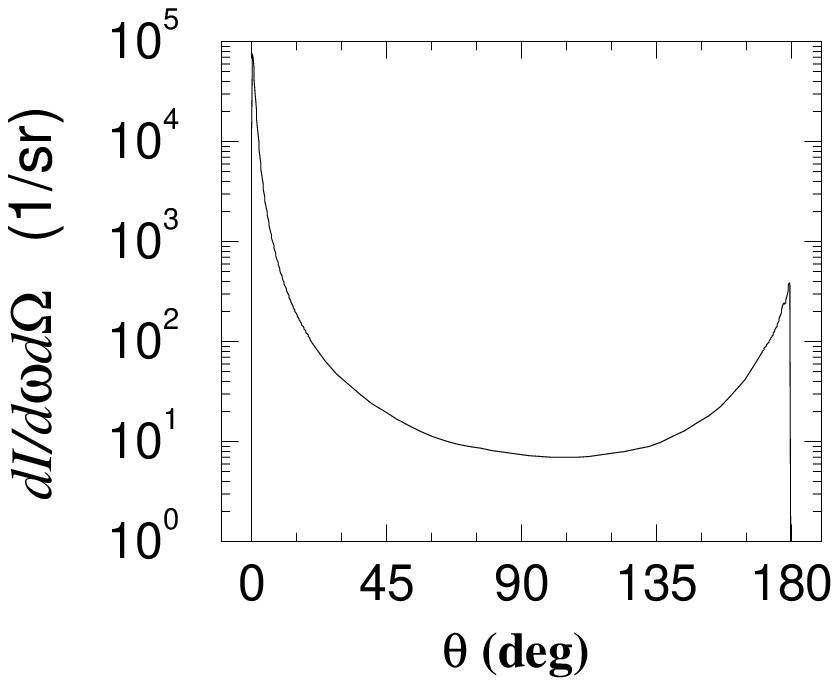,width=3.50in}}
\caption{Highly asymmetric photon emission from collisions of an 
artificial 197 proton nucleus with an artificial 197 neutron nucleus.} 
\label{f:pnth}
\efi
\bfi
\centerline{\epsfig{figure=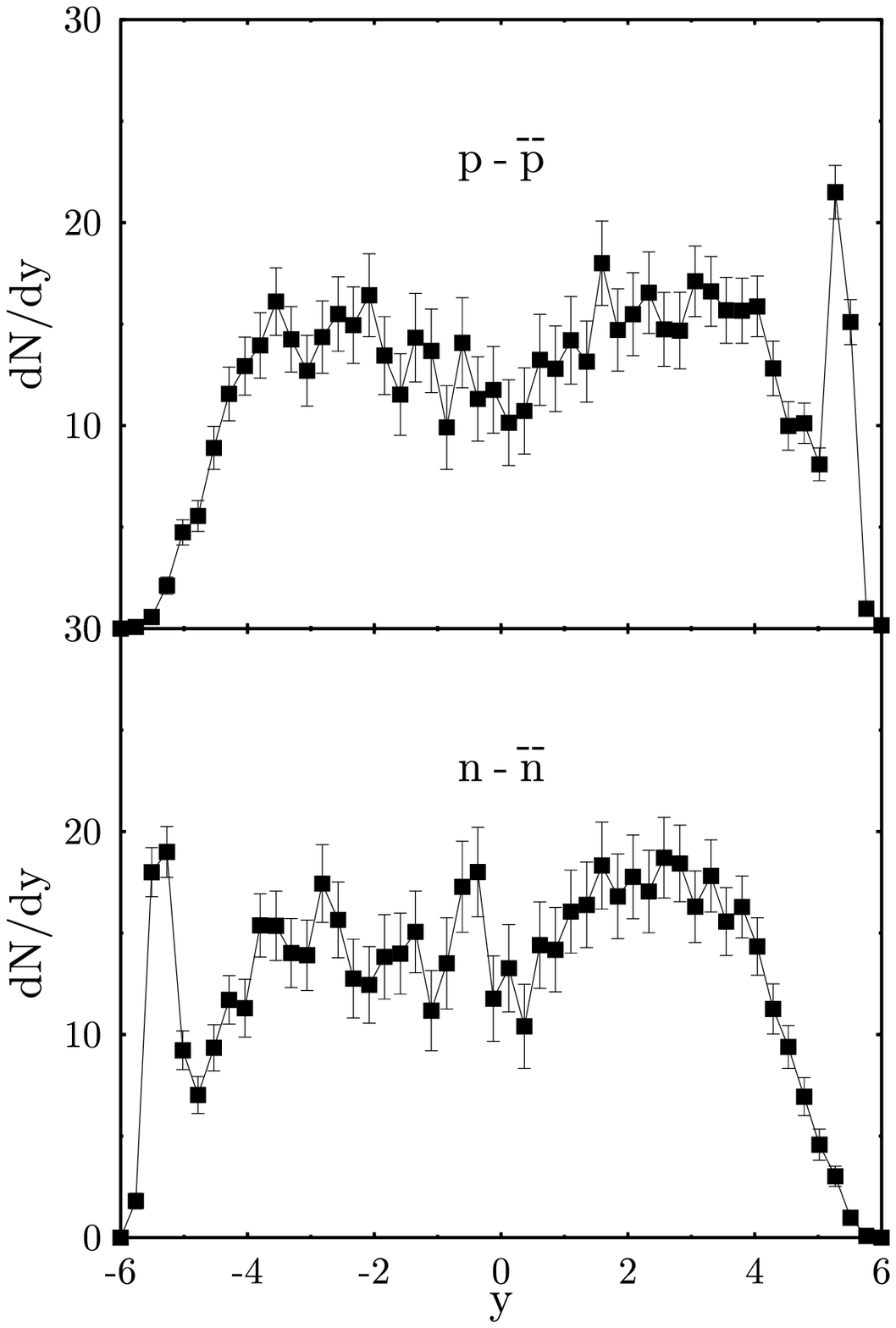,width=3.0in}}
\caption{The final net rapidity distribution of protons and neutrons
from the collisions of artificial nuclei. Except for the peaks
from the remnants of the initial nuclei, the nucleons spread
out quite evenly over the entire rapidity range.} 
\label{f:pnrap}
\efi

\subsection{Time dependence} 
\label{s:time} 

Another question of interest is the time scale of nucleus-nucleus collisions.
There is a paucity of direct means of getting any estimates and a paucity of 
meaningful observables.  One might think of Hanbury-Brown and Twiss hadron 
interferometry, but this only relates to the time interval over which the 
final state hadrons emerge. In \cite{kw} the possibility of using 
bremsstrahlung for this purpose was discussed but that, in part, relied 
on the collisions being Landau-like.  Theoretically we can examine the 
time-dependence of the emitted bremsstrahlung. In \fref{f:pc}, the ratio
of the instantaneous to final intensity has been plotted at two extreme 
values of $\q$ as a function of time. It shows that at small angle, 
for example at $\q=1^o$, most of the soft bremsstrahlung were emitted 
within the first 5 fm/c. This results from the fact that in the longitudinal 
direction the bulk of the charges have finished their acceleration within a 
very brief duration upon first contact. In the transverse direction,
on the other hand, it takes much longer, something like 100 fm/c, 
for 90\% of the soft photons to be emitted. This is logical since
we expect the transverse expansion to become significant only at the
later stages. These facts here are essentially energy independent or 
weakly dependent in the range $\o < 200$ MeV as seen in \fref{f:om}. 

\bfi
\centerline{\epsfig{figure=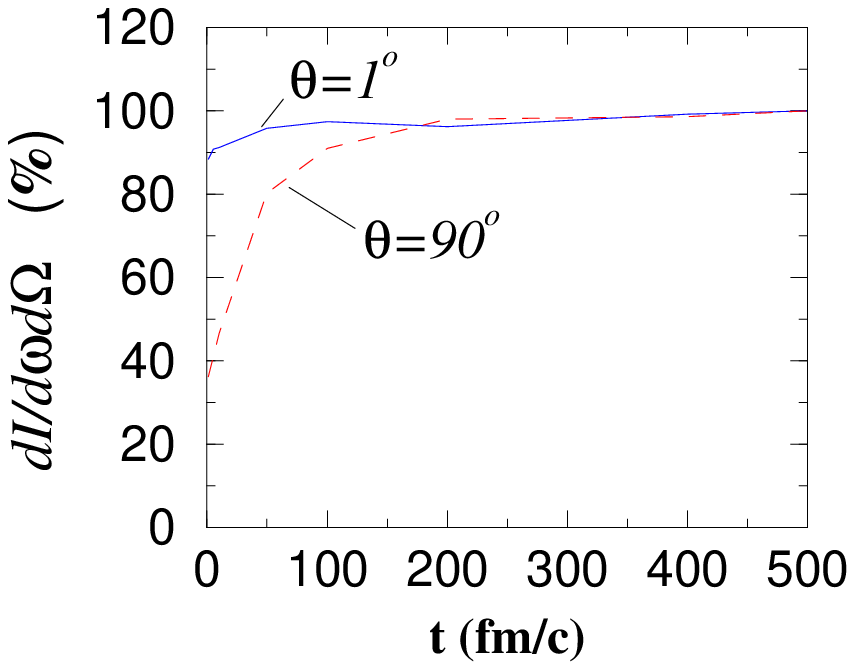,width=3.50in}}
\caption{The percentage of the bremsstrahlung final intensity at
two values of $\q$ as a function of time. Solid line is result
at $\q=1^o$ and dashed line for $\q=90^o$.  Here $\omega = 10$ MeV.} 
\label{f:pc}
\efi

\subsection{Global Charge Stopping Time}
\label{s:tf}

In Sect. \ref{s:qo}, we have seen that there are signs of 
oscillation in the intensity versus energy plots in \fref{f:om} 
and \fref{f:nvp} which revealed a partial Landau-like collision
picture. In \cite{kw} the Landau-like scenario was found to
have one advantage over the Bjorken-like scenario, which was that 
the amplitude and frequency of the oscillation was sensitive to
the global charge stopping time $t_f$. We now attempt to extract 
$t_f$ from the UrQMD data using the simple Landau-like model in \cite{kw}. 
In that paper a flat rapidity distribution, or using the defined
quantity from \cite{sw} $S=1/2$, was assumed throughout. We will see 
below that UrQMD data yields somewhat more than one-half stopping, 
or $S > 1/2$. Therefore we have to adjust the intensity distribution
from the simple model by a constant upward shift in order to 
fit the data. At smaller angles the oscillation is either unclear or its
amplitude too small to be of practical use. So it is at $\q=20^o$ that 
the adjusted intensity from the model will be fitted to that of UrQMD.
The result is shown in \fref{f:land}. It is seen that the range
1.2 $<t_f<$ 1.4 fm/c essentially covers the data. It can therefore be 
deduced that the bulk of the charges settled extremely quickly.  
\bfi
\centerline{\epsfig{figure=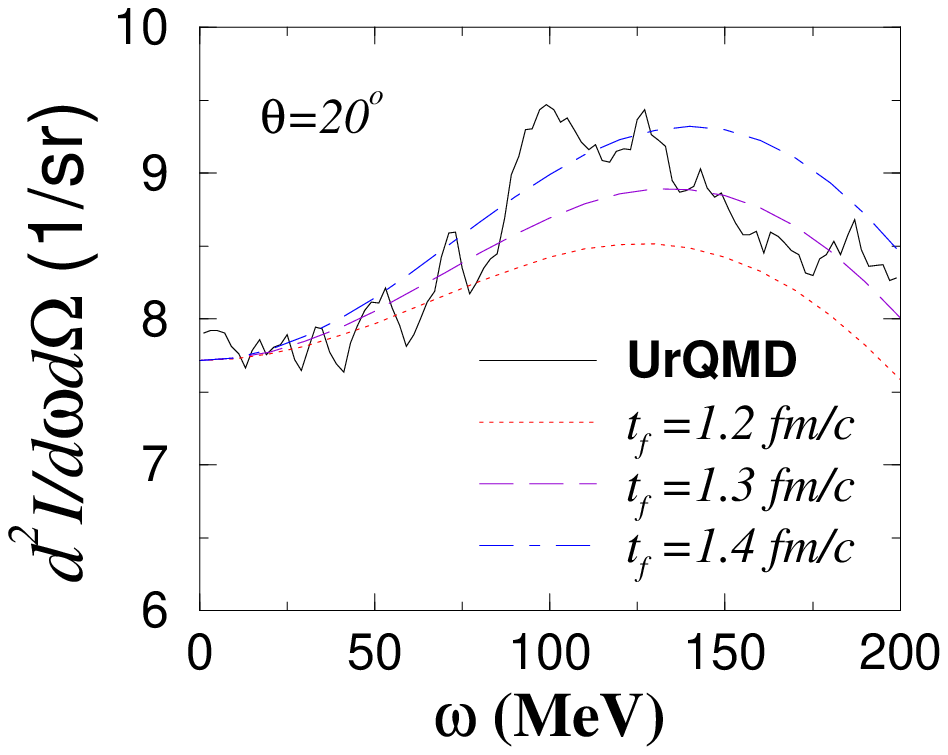,width=3.50in}}
\caption{The simple Landau-like model from \protect\cite{kw} is fitted 
to the UrQMD data with three values of $t_f$.}  
\label{f:land}
\efi
To see if this is indeed the case, snapshots of the net-baryon and 
net-charge rapidity distributions have been taken from UrQMD at several 
times. These are shown in \fref{f:snaps}. We see that these rapidity
distributions settled already at times around 1 fm/c and do not change
much after that\footnote{The net-charge distribution shows 
larger fluctuations at $t=20$ fm/c than at previous times. This
may be due to resonance decays into mesons.}. The global stopping time 
determined above, although very short, is therefore sensible. 
The point is that the value of $t_f$ obtained from bremsstrahlung 
agrees with the time it takes for the baryon and electric charge
rapidity distributions to acquire their final form. 
We have now shown that $t_f$ can be determined from bremsstrahlung, 
a piece of information that no hadronic measurement could provide. 

\bfi
\centerline{\epsfig{figure=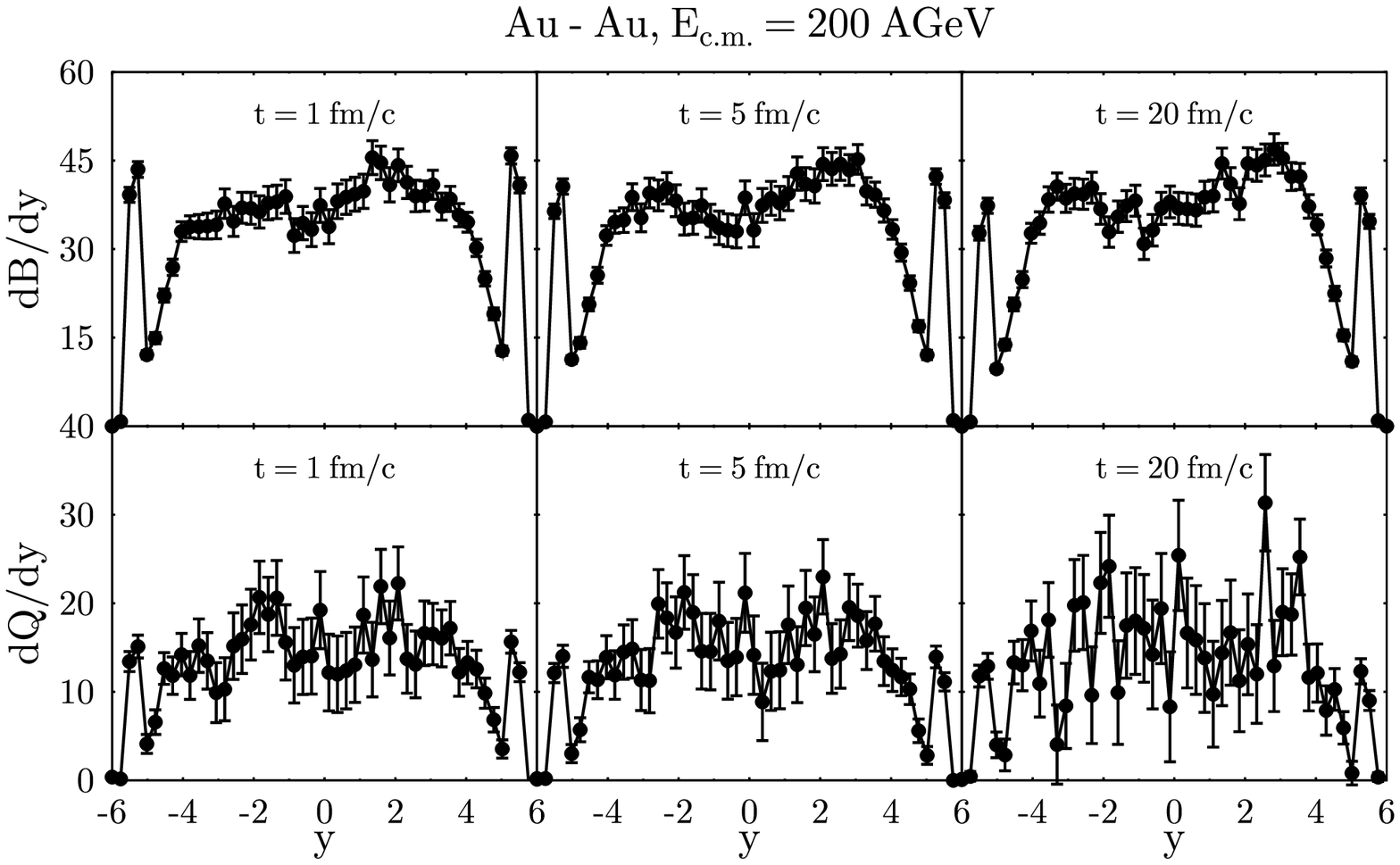,width=7.0in}}
\caption{Net-baryon and net-charge rapidity distribution from UrQMD 
at various times.}
\label{f:snaps}
\efi

\section{Inference of Charge vs. Baryon Stopping} 
\label{s:mtm}

In an ideal world for measuring stopping, bremsstrahlung would 
be emitted only from charged baryons and anti-baryons and the 
contribution from the many produced mesons, which can be both positively 
as well as negatively charged, would cancel each other. For Au+Au 
collisions, the largest contributions do indeed come from baryons 
and anti-baryons but there remains a non-negligible contribution from 
mesons, as discussed in Sect. \ref{s:qo}. The mesonic contribution has 
to be subtracted somehow if bremsstrahlung is to be used to probe 
baryon stopping, as opposed to charge stopping. To do this, we have 
to understand from the soft photons' point of view the essential 
dynamics of the sources. We find it illuminating and instructive 
to use another, much simpler, model that can reproduce, as far as 
bremsstrahlung is concerned, the generated data from UrQMD. In this 
section we will pretend that the bremsstrahlung from UrQMD is 
experimental data and show that the procedure described below is 
capable of extracting the degree of stopping in UrQMD. 

Several facts from UrQMD prove to be useful in this regard. 
We notice that for soft photon emission there is a fundamental 
difference between the contributions from baryons and from mesons. For 
the emission from baryons and anti-baryons the precise values of their 
transverse velocities (relative to the beam direction) are unimportant. 
One could set these to zero without affecting bremsstrahlung very much at
all; recall \fref{f:nvp}. On the contrary, transverse acceleration of the 
mesons is very important.  We could therefore model the radiation from 
UrQMD by assuming that baryons and anti-baryons move only longitudinally 
with no transverse velocity. Information on the charged baryons 
and anti-baryons for soft photon emission is therefore all encoded in the 
final rapidity distributions. This is, of course, not sufficient as
we have already mentioned that mesons contribute a finite amount
to the radiation.  In some sense the mesons are ``noise'' in this context and 
their transverse velocities are an important part of it. We have found that 
the ``noise'' can be reasonably subtracted by modeling the motion of 
the mesons in the manner to be described below. 

In \cite{sw} it was suggested to measure low energy photons
at the special angle $\qh \simeq 7.93^o$ 
at RHIC to determine the amount of stopping via the stopping variable $\cs$. 
\be
\cs = 1 - \frac{(1-v_0^2 \cos^2\qh)} {2v_0^2 \cos\qh} 
      \int_{-\infty}^{+\infty} dy\,\frac{v(y)\rho(y)}{1-v(y)\cos\qh}
\ee
Here $v(y)=\tanh(y)$ is the velocity along the beam axis and $\rho(y)$ is 
the baryon rapidity distribution normalized to 2 for collisions between 
symmetric nuclei.  $\cs$ ranges from 0 for no stopping to 1 for full 
stopping.  The angle $\qh$ depends on the beam velocity $v_0$ and is 
chosen so that $\cs = 1/2$ for any distribution that belongs intuitively 
and obviously to one-half stopping, such as 
$\r(y) \sim \d(y-y_0)+\d(y+y_0)+2\d(y)$ or $\r(y)=$ constant.  
Defining baryon stopping in terms of the above variable permits a direct 
connection between soft bremsstrahlung and baryon stopping, as discussed 
in \cite{sw}. 
  
In the soft photon limit, due to destructive interference,
only the initial and final charges contribute to the bremsstrahlung. 
This permits a well-known simplification. That is, the 
detailed collision history can be ignored and only the initial and final 
particles need to be taken into account. In the presence of the above 
mentioned ``noise'' the relation between $\cs$ given above and the 
soft photon intensity distribution needs to be modified. 
Without loss of generality, we choose the directional vector 
$\N$ to be $\N =(\sin \q, 0, \cos \q)$. We define
\be \M = \sum_{i \in {\text{mesons}}} \frac{q_i\;\V_i}{1-\N\cdot \V_i} 
\label{eq:m}
\ee
for the sum over products of the charge and velocity factor over 
the final mesons.  With the aforementioned mesonic ``noise'' included 
we have to modify the relation between $\cs$ and the bremsstrahlung
intensity distribution $dI/d\o d\Omega$ at the angle $\q_{1/2}$. 
The required modification amounts to simply replacing $\cs$ as 
appeared in Eq. (13) of ref. \cite{sw}, which is repeated here for
convenience 
\be \cs = \frac{1-v_0^2 \cos^2 \qh}{v_0^2 \;\sin 2\,\qh} \;
    \Big ( \frac{4\p^2}{\a Z^2} \;
    \frac{d^2 I}{d\o\, d\Omega} \Big |_{{\o \ra 0\;\;\;\;} \atop {\q =\qh}} 
    \Big )^{1/2}      \; ,
\label{eq:bd}
\ee
by 
\be \cs \lra \left \{
       \left (\cs 
     - \frac{(1-v_0^2 \cos^2\qh)(M_z -M_x \cot \qh)}{2 Z v_0^2 \cos\qh} 
       \right )^2
     + \frac{(1-v_0^2 \cos^2\qh)^2}{4 Z^2 v_0^4 \cos^2 \qh} M_y^2 
             \right \}^{1/2} \, .
\label{eq:s_repl}
\ee
Here $v_0 \simeq 0.999956$ at RHIC.

With the above replacement, the procedure to get 
$\cs$ is less straightforward but nevertheless possible. 
We find that the final meson contribution to soft photon emission 
is consistent with that of $N_+$ positively and $N_-$ negatively 
charged particles emerging from the collisions, apart from 
the expected forward-backward bias, in random directions. 
With this observation we arrive at the following 
procedure to obtain an estimate of $\cs$.  

\begin{itemize}

\item[(1)]{$N_+$ and $N_-$, which represent the approximate 
numbers of the corresponding charged mesons, are generated randomly
within a given range. Models such as UrQMD 
can provide reasonable values; for example, on average 
$\lan N_+ \ran \simeq \lan N_- \ran \simeq 3000$ at RHIC.}

\item[(2)]{The difference $\D Q = N_+ -N_-$, where 
$\D Q \ll N_+, N_-$, need not be equal to zero but is bounded within 
the range $0 \leqslant \D Q < 2 Z$. In fact, as seen in experiment 
at SPS and from the UrQMD model, the average value of $\D Q$ is in 
the lower end of the range below 40. The upper limit is only a
statistically improbable theoretical possibility.} 

\item[(3)]{Equal numbers of these charges will have positive and
negative longitudinal velocities. Apart from that rapidities are 
randomly assigned according to the triangular distribution 
shown in \fref{f:tri} (this is the simplest yet reasonable 
distribution which agrees fairly well with realistic distributions). 
This gives a sensible forward-backward bias.} 

\bfi
\setlength{\unitlength}{1.0mm}
\begin{center}
\begin{picture}(60,27)
\put(5,5){\line(1,0){50}}
\put(30,2){\line(0,1){25}}
\put(7,5){\line(0,1){2}}
\put(53,5){\line(0,1){2}}
\put(3,2){$-y_0$}
\put(52,2){$y_0$}
\put(7,5){\line(3,2){23}} 
\put(53,5){\line(-3,2){23}}
\put(31,22){$1/y_0$} 
\put(29,-1){$y$}
\end{picture}
\end{center}
\caption{}
\label{f:tri}
\efi

\item[(4)]{The velocities of the majority of the charges are close
to the speed of light. Slower longitudinally moving charges 
contribute negligibly to the bremsstrahlung because of the 
denominator in \eref{eq:m}. For these charges, the transverse 
component of $M_x$ and $M_y$ are not negligible. For the purpose of 
giving transverse velocities to our simulated mesonic charges,
a maximum transverse velocity $\vm$ close to the initial speed $v_0$
is necessary to give reasonable results. Note 
that this limit on the velocities is for $v_x$ and $v_y$. It does not 
affect $v_z$, which is distributed according to the discussion in 
point (3) above.} 

\item[(5)]{The direction of the transverse motion of each
individual charged particle is picked randomly subject to the
constraint that the total sum of the transverse velocities is
approximately zero.} 

\item[(6)]{With the above steps in setting the velocities, directions
and numbers of the mesonic charges, the 3-vector $\M$ and therefore
the intensity distribution are calculated over a number of events. 
Using the relation between \eref{eq:s_repl} and the intensity distribution 
given in \cite{sw}, the mean $dI/d\o d\Omega$ in the soft photon limit
and the standard deviation $\s$ of the intensity generated from 
UrQMD can be recovered by suitable choice of the value of $\cs$.} 

\item[(7)]{When experimental information becomes available, one simply replaces 
the intensity from the UrQMD model at the angle $\qh$ with experimental 
measurements and repeat the above procedure to obtain $\cs$.} 

\end{itemize}

For Au+Au at RHIC 
and at the angle $\qh$ from the beam direction the intensity distribution 
and its standard deviation of soft photons from the UrQMD model are, 
respectively 
\be  \frac{dI}{d\o d\Omega} \Big |_{{\o \ra 0\;\;\;\;} \atop {\q =\qh}}
     = 94.1  {\text{\ \ \ and \ \ \ }} 
     \s = 71.8   \; .
\label{eq:mr}
\ee
(Note the large intrinsic fluctuation in the intensity; this is
characteristic of soft photons as computed in UrQMD.) 
The other relevant quantities from the model are
\be \adq \simeq 19 {\text{\ \ \ and \ \ \ }} \cs = 0.547 \; .
\label{eq:ms}
\ee
The latter means that there is just over half stopping \cite{sw}
in the collisions predicted by UrQMD at RHIC. In \fref{f:aurap},
the net-proton and net-neutron rapidity distribution from Au+Au
at RHIC are plotted. They show exactly what the quantity $\cs$
is designed to indicate, which is there is just over half stopping. 
The standard deviation of the intensity, $\s$, is quite large, indicating very 
large fluctuations in the charge rapidity distribution on an event by event 
basis.  There are no such fluctuations in hydrodynamic treatments of heavy ion 
collisions, of course. 

To the parameters already introduced, we now add $\ci$ 
\be \ci(\q) = \int^{+\infty}_{-\infty} dy \frac{v(y)\, \r(y)}{1-v(y) \cos \q}\;.
\ee
In terms of $\ci$ the baryon stopping parameter $\cs$ is 
\be \cs = 1 -\frac{(1-v_0^2 \cos^2 \qh)}{2 v_0^2 \cos \qh} \; \ci(\qh) \; .
\ee
We can now vary them to see if it is possible to recover the results from UrQMD. 

\bfi
\centerline{\epsfig{figure=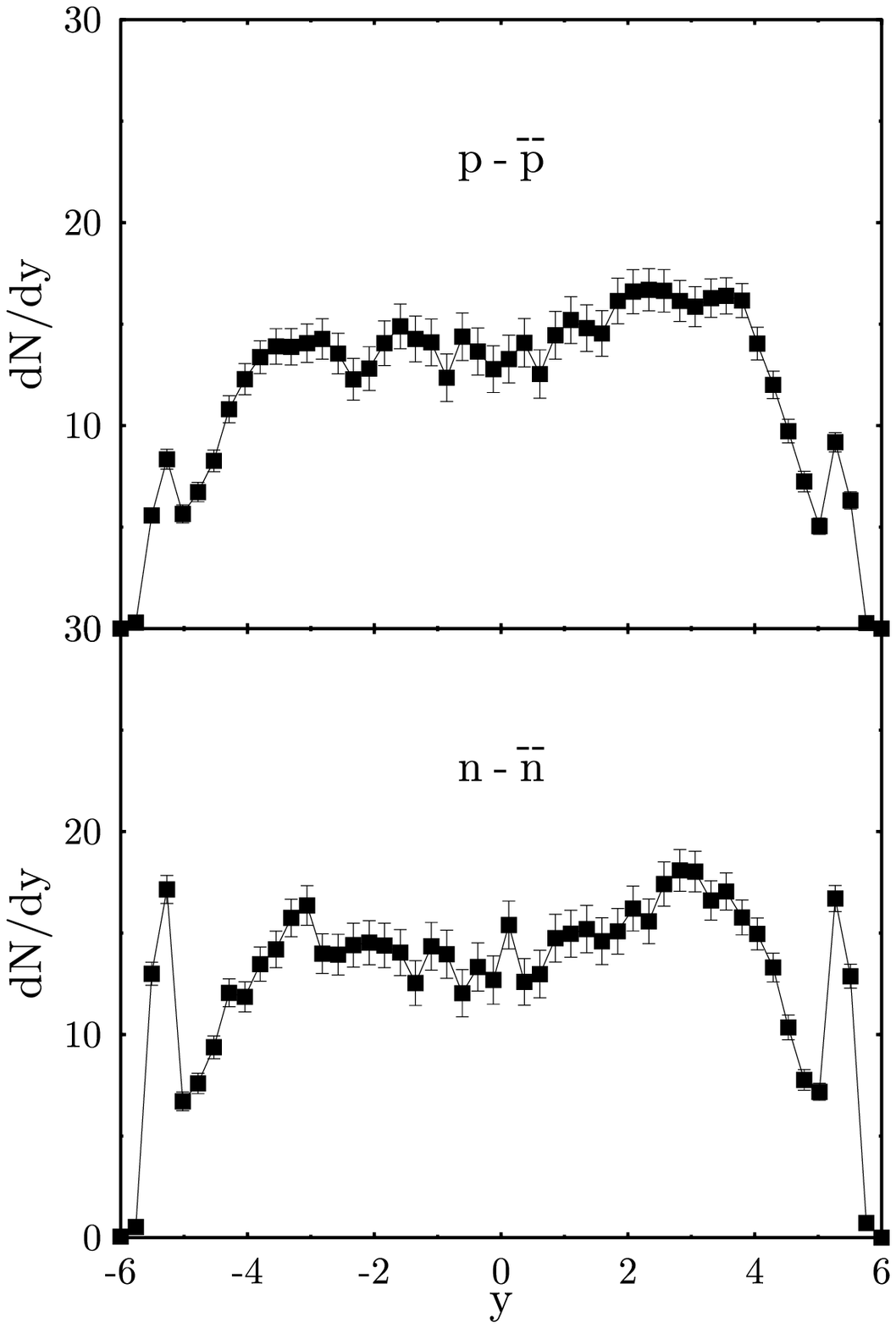,width=3.0in}}
\caption{Final net-proton and net-neutron rapidity distribution from 
Au+Au collisions at RHIC.} 
\label{f:aurap}
\efi

After some trial and error, certain facts can be established.
Of all the parameters the intensity is most sensitive to $\cs$, 
which is good. The next most important parameter is $\adq$. 
The rest are for fine tuning the results. These are tabulated 
in \tref{t:s} with various input parameters. Initially setting 
both $\adq$ and $\ci$ to zero gives a result that is too large 
(see the first row of the table). $\vm$ is temporarily set to 
0.997 here for exploratory purpose. Next, increasing $\adq$ to 
the largest but very unlikely value of $2 Z$ is still not 
sufficient to reduce the intensity to the actual value. This 
is shown in the second row of the table. On the other hand, by 
setting $\adq =0$ and increasing $\ci$ to about 48.5 reasonable 
agreement can be obtained. From experiments at SPS, from models, 
and from very general considerations, we expect that some of the 
initial positive charge will be transferred to the mesons. 
Reasonable values for $\adq$ are within the range 10--30. 
This range can help us get a bound on the value of $\cs$ because 
once a fit has been established for, say $\adq= 10$, increasing 
this to 30 will require a corresponding decrease in $\ci$ to 
compensate. 

The next question is what value should $\vm$ be used. Fortunately, 
it turns out that $\cs$ is not very sensitive to $\vm$. 
To see this, we picked the angle $\q=5^o$. From UrQMD
we know the following results.
\be  \frac{dI}{d\o d\Omega} \Big |_{{\o \ra 0\;\;\;\;} \atop {\q =5^o}}
     = 280  {\text{\ \ \ and \ \ \ }} 
     \s = 171
\label{eq:5}
\ee 
Also the value $\ci$ at this angle is $\ci(5^o) = 95.0$. 
In the next four rows in the table, it is shown that only with 
$\vm =0.999$ or higher is this value of $\ci$ able to yield 
intensities that can agree with that from the model. Below this value, 
the resulting intensities are just short. $\cs$ is only defined and 
meaningful at $\qh$ so the $\cs$ column was left blank.
Although in real experiments one does not know {\it a priori} the value of 
$\ci$, the remaining entries in \tref{t:s} should be enough to show 
that a precise knowledge of the value of $\vm$ is unnecessary. 
The range of values of $\vm$ from 0.997 to 0.999 and higher 
produce a range of $\ci$ that covers the correct value. 
Here the covered range based on the $\adq =$10 and 30 limit for
$\vm =$0.997 and 0.999 are shown. They are sufficiently narrow to
convey the fact that baryon stopping is somewhat over halfway,
exactly what has been plotted in \fref{f:aurap}. 

\begin{minipage}[h]{\columnwidth}
\begin{table}
\caption{Various parameters used for fitting $\cs$ to the intensity
distribution from UrQMD.}
\label{t:s}
\begin{tabular}{cccccccc} 
 $\lan N_+ \ran $ & $\adq$ & $\vm$  & $\q$ & $\ci$ 
 & $\cs$ & $\frac{dI}{d\o d\Omega}$ & $\s$ \vspace{1mm} \\  \hline\hline
 2929 &   0 & 0.997 & $\qh$ &  0.0 & 1.0    & 269.2 & 120.2 \\ 
 2929 & 158 & 0.997 & $\qh$ &  0.0 & 1.0    & 146.1 &  94.3 \\
 2987 &   0 & 0.997 & $\qh$ & 48.5 & 0.532 &  95.6 &  76.7 \\ \hline\hline
 3032 &  10 & 0.997 & $5^o$ & 95.0 & NA     & 273.9 & 165.5 \\
 2969 &  10 & 0.998 & $5^o$ & 95.0 & NA     & 274.0 & 178.0 \\
 3009 &  10 & 0.999 & $5^o$ & 96.5 & NA     & 280.7 & 184.5 \\ 
 2979 &  30 & 0.999 & $5^o$ & 91.0 & NA     & 280.5 & 160.7 \\ \hline\hline
 2978 &  10 & 0.999 & $\qh$ & 48.0 & 0.537 &  94.1 &  76.1 \\
 3002 &  30 & 0.999 & $\qh$ & 45.9 & 0.557 &  94.3 &  78.3 \\ \hline\hline
 2978 &  10 & 0.997 & $\qh$ & 47.5 & 0.542 &  94.4 &  76.2 \\ 
 2994 &  30 & 0.997 & $\qh$ & 43.0 & 0.585 &  95.8 &  71.7 \\ 
\end{tabular} 

\end{table}
\end{minipage}

Our procedure yields the value of stopping around $\cs =$ 0.537 
-- 0.585, which is close enough to the value 0.547 from UrQMD in 
\eref{eq:ms}. So, with our extraction procedure to separate the charged 
baryon and charged meson contribution, soft bremsstrahlung 
from nuclear collisions can definitely be used to probe the degree 
of stopping. The procedure described here is relatively simple for
subtracting the charged meson contributions. When experimental data
are available eventually, one could of course use other more complex 
models such as UrQMD itself for the subtraction but the advantages of 
our procedure are that it is much simpler, more efficient therefore 
can be done by almost anyone using a very small program under fifty
lines of codes.

\section{Conclusion} 

In this paper we have used the microscopic, dynamical model UrQMD to compute 
bremsstrahlung emitted by baryons and mesons as they are created, annihilated, 
and scattered during a central collision between Au nuclei at RHIC.  Fine grained 
details on time and length scales shorter than about 1 fm do not matter from the 
point of view of emission of photons of energy less than about 200 MeV. 
Therefore, as discussed in earlier papers and demonstrated here with a specific 
microscopic model, bremsstrahlung measurements can probe the space-time 
evolution of high energy heavy ion collisions.  UrQMD suggests that electric 
charge easily diffuses in rapidity space, giving rise to an approximately flat 
charged rapidity distribution.  This was concretely demonstrated by colliding 
pure proton nuclei on pure neutron nuclei in the computer.  
We also examined the connection between 
charge stopping and baryon stopping, and showed that the noise caused by net 
charge transfer from protons to mesons can be simulated sufficiently accurately 
so that the two measures of stopping can still be related.  In particular, the 
bremsstrahlung intensity at small angles approximately scales with the degree of 
stopping as quantified in the variable $\cs$.

Bremsstrahlung is an interesting and useful way to study the dynamics of high
Energy heavy ion collisions.
Unfortunately, it is unlikely than the initial round of experiments at RHIC will 
be able to measure photons of sufficiently low energy as to be useful in this 
context.  Fortunately, there are still two collision halls vacant at RHIC which 
could accommodate a dedicated soft photon detector.  Future experiments should
take advantage of these unique possibilities.

\section*{Acknowledgments}

This work was supported by the U.S. Department of Energy under 
grant no. DE-FG02-87ER40328. S.A.B was supported by the National 
Science Foundation, grant no. PHY-00-70818 and M.B. received support 
by the A. v. Humboldt Foundation.



\begin{thebibliography}{99}


\bibitem{qm} See the proceedings of the Quark Matter conferences, the most
recent in print being: Nucl. Phys. {\bf A661}, (1999).

\bibitem{Landau} L. D. Landau, Izv. Akad. Nauk SSSR (Physics Series) {\bf 17},
51 (1953);
S. Z. Belenkij and L. D. Landau, Uspekhi Fiz. Nauk, {\bf 56}, 
309 (1955);
Nuovo Cimento Suppl. {\bf 3}, 15 (1956).

\bibitem{Bj}J.D. Bjorken, \jou{\PRD}{27}{140}{1983}. 

\bibitem{wienold96a}T. Wienold et al. \jou{\NPA}{610}{76}{1996}.

\bibitem{na49}NA49 collaboration, \jou{\PRL}{82}{2471}{1999}. 

\bibitem{kap} J. Kapusta, Phys. Rev. C {\bf 15}, 1580 (1977).

\bibitem{BM}
J. D. Bjorken and L. McLerran, Phys. Rev. D {\bf 31}, 63 (1985).

\bibitem{Dumitru}
A. Dumitru, L. McLerran, H. St\"ocker and W. Greiner, Phys. Lett.
B {\bf 318}, 583 (1993).

\bibitem{jkcs}S. Jeon, J. Kapusta, A. Chikanian and J. Sandweiss, 
\jou{\PRC}{58}{1666}{1998}.

\bibitem{kstalk}J.I. Kapusta and S.M.H. Wong, {\it Proceedings of the 
XXIX International Symposium on Multiparticle Dynamics}, August 1999, 
edited by I. Sarcevic and C.-I. Tan (World Scientific Singapore, 2000), 
p. 346 [hep-ph/9909573]. 

\bibitem{kw}J.I. Kapusta and S.M.H. Wong, \jou{\PRC}{59}{3317}{1999}.

\bibitem{eesg} U. Eichmann, C. Ernst, L.M. Saratov, and W. Greiner,
\jou{\PRC}{62}{044902}{2000}.

\bibitem{vmg}D. Vasak, B. M\"uller and W. Greiner, \jou{\JPG}{11}{1309}{1985}.

\bibitem{sumgv}T. Stahl, M. Uhlig, B. M\"uller, W. Greiner, D. Vasak, 
\jou{\ZPA}{327}{311}{1987}.

\bibitem{hmsg}R. Heuer, B. M\"uller, H. St\"ocker, and W. Greiner,
\jou{\ZPA}{330}{315}{1988}.

\bibitem{urqmd1} 
S.~A.~Bass, M.~Belkacem, M.~Bleicher, M.~Brandstetter, L. Bravina, C.~Ernst, 
L.~Gerland,  M.~Hofmann, S.~Hofmann, J.~Konopka, G.~Mao, L.~Neise, S.~Soff,
C.~Spieles, H.~Weber, L. A. Winckelmann, H. St\"ocker, W. Greiner, 
Ch.~Hartnack, J. Aichelin and N. Amelin,
Progr. Part. Nucl. Phys. {\bf 41}, 255 (1998).

\bibitem{urqmd2}
M. Bleicher, E. Zabrodin, C. Spieles, S. A. Bass, C. Ernst, S. Soff, L. 
Bravina, M. Belkacem, H. Weber, H. St\"ocker and  W. Greiner, J. Phys. G: 
Nucl. Part. Phys. {\bf 25}, 1859 (1999).

\bibitem{sw}S.M.H. Wong, \jou{\PLB}{480}{65}{2000}. 

\bibitem{pdb} 
Particle Data Group, \jou{\EPJC}{3}{1}{1998}.

\bibitem{js}J. D. Jackson, {\it Classical Electrodynamics}, 
2nd ed. (Wiley, N.Y. 1975).  


\end{thebibliography}
\end{document}